\documentclass[aps,prd,twocolumn,showpacs,preprintnumbers,nofootinbib,amsmath,amssymb]{revtex4}

\usepackage{graphicx}
\usepackage{dcolumn}
\usepackage{bm}
\usepackage{amsmath}

\newcommand{\beq}{\begin{eqnarray}}
\newcommand{\eeq}{\end{eqnarray}}

\newcommand{\real}{{\sf I}\kern-.12em{\sf R}}
\newcommand{\comp}{{\sf I}\kern-.50em{\sf C}}
\newcommand{\unity}{{\sf I}\kern-.54em{\sf 1}}

\def\spose#1{\hbox to 0pt{#1\hss}}
\def\ltapprox{\mathrel{\spose{\lower 3pt\hbox{$\mathchar"218$}}
 \raise 2.0pt\hbox{$\mathchar"13C$}}}

\begin{document}

\title{QCD Phase Transition in a Strong Magnetic Background}
\author{Massimo D'Elia$^1$, Swagato Mukherjee$^2$ and Francesco Sanfilippo$^3$}
\affiliation{$^1$Dipartimento di Fisica, Universit\`a
di Genova and INFN - Sezione di Genova, Via Dodecaneso 33, 16146 Genova, Italy\\
$^2$Physics Department, Brookhaven National Laboratory, Upton, NY 11973-5000, USA\\
$^3$Dipartimento di Fisica, 
Universit\`a di Roma ``La Sapienza'' and INFN - Sezione di Roma, P.le A. Moro 5, 00185 Roma, Italy}
\date{\today}

\begin{abstract}
We investigate the properties of the deconfining/chiral restoring transition for two
flavor QCD in presence of a uniform background magnetic field.  We adopt standard
staggered fermions and a lattice spacing of the order of
0.3 fm. We explore different values of the bare quark mass, corresponding to pion
masses in the range 200 - 480 MeV, and magnetic fields up to $|e|B\sim0.75$ GeV$^2$.
The deconfinement and chiral symmetry restoration temperatures remain compatible with
each other and rise very slightly ($<2\%$ for our largest magnetic field) as a
function of the magnetic field. On the other hand, the transition seems to become
sharper as the magnetic field increases. 
\end{abstract}

\maketitle

\section{Introduction}

The study of strong interactions in presence of magnetic background fields is
relevant to many phenomenological contexts.  Large magnetic fields ($B \sim 10^{16}$ Tesla, i.e. $\sqrt{|e|B} \sim 1.5$ GeV) may have been produced at the
cosmological electroweak phase transition~\cite{Vachaspati:1991nm} and they may have
influenced subsequent QCD transitions.  Slightly lower fields are expected to be produced in non-central heavy
ion collisions, reaching up to $10^{14}$ Tesla at RHIC and up to $\sim  10^{15}$
Tesla at LHC~\cite{heavyionfield1,heavyionfield2}. Large magnetic fields, of the
order of $10^{10}$ Tesla, are also expected in some neutron stars known as
magnetars~\cite{magnetars}.

The influence of electric and magnetic fields on the chiral properties of the vacuum
has been studied since some time, using various approximations or effective models of
QCD~\cite{Klevansky:1989vi,Suganuma,Gusynin:1994xp,Shushpanov:1997sf,Ebert,Miransky:2002rp,Cohen:2007bt,Zayakin:2008cy},
predicting an enhancement of chiral symmetry breaking as
a magnetic field is switched on.  Recently new interesting phenomenology has
been proposed, consisting in the appearance of an electric current parallel to the
magnetic field in presence of deconfined quarks and local CP violations, 
induced e.g. by topological charge fluctuations~\cite{cme0,cme1}.  That is
usually known as the chiral magnetic effect and experimental confirmations of it are
currently being searched by heavy ion experiments~\cite{star}.

An important issue is the influence of the
magnetic field on the structure of the QCD phase diagram, in particular on the
location and the nature of deconfinement and chiral symmetry restoration. Clarifying
that in the case of strong magnetic fields is essential to correctly predict the
phenomenological consequences of the QCD transition on the evolution of the Universe
during its early stages.  Some computations exist, based on different
approximations and QCD-like 
models~\cite{Klimenko,agasian,fraga1,NJL,fukushima,fraga2,gatto}, which predict the 
possibility of a quite rich phenomenology, ranging from a possible splitting of
deconfinement and chiral symmetry restoration to a sizable increase in the strength
of the transition.  However the various model predictions are not always consistent
among themselves. 

A clarification of these issues may come from lattice QCD
computations.  A magnetic background field, contrary to an electric field or a finite baryon density,
does not give rise to technical difficulties 
such as a sign problem.
The phase diagram in presence of
a chromo-magnetic background field has been investigated in Refs.~\cite{bari1,bari2}: 
the transition temperature decreases as a function of
the external field, with deconfinement and chiral symmetry restoration
staying strictly related to each other. Investigations in presence of
(electro-) magnetic fields have been done since long with the purpose
of studying the magnetic properties of hadrons~\cite{hadron1,hadron2}, while some recent
studies~\cite{cme-itep,cme-blum,itep1,kharzeev-itep} have reported mostly on the chiral
properties of the theory and about numerical evidence for the chiral magnetic effect.

In this paper we report on a first investigation of the QCD phase
transition in presence of an (electro-) magnetic background field. In order to do that,
it is essential to include dynamical quark contributions, since only quark fields, being electrically
charged, are influenced by the magnetic field. We have
considered $N_f = 2$ QCD with standard staggered fermions and
different values of the quark masses, to appreciate how the effects of the
magnetic field change as the mass spectrum changes (in the heavy quark limit the
magnetic field becomes irrelevant). In Section~\ref{setup} we give some details
about lattice QCD in presence of a background field and about our
numerical setup. In Section~\ref{results} we present our numerical results and
finally, in Section~\ref{conclusions}, we give our conclusions.

\section{Numerical Setup}
\label{setup}

We consider $N_f = 2$ QCD, with quarks carrying different electric
charges and coupled to a background (electro) magnetic (e.m.) field.  
The background field affects quark propagation and
corresponds to a modification of the Dirac operator. In the continuum the
covariant derivative changes by inclusion of the e.m. $A_\mu$ field; on
the lattice one has to add appropriate $U(1)$ fields to the gauge link variables
which parallel transport quarks fields from one
lattice site to the other. In the case of a uniform magnetic field $B$,
with different electric charges for the two flavors, $q_u = 2|e|/3$ and
$q_d = -|e|/3$ ($|e|$ being the elementary charge), the partition function
of the (rooted) staggered fermion discretized version of the theory is
\beq
Z(T,B) \equiv \int \mathcal{D}U e^{-S_{G}} 
\det M^{1\over 4} [B,q_u]
\det M^{1\over 4} [B,q_d]
\:,
\label{partfun1}
\eeq
\begin{eqnarray}
M_{i,j} [B,q] &=& a m \delta_{i,j} 
+ {1 \over 2} \sum_{\nu=1}^{4}\eta_{i,\nu} \left(
\vphantom{ U^{\dag}_{i-\hat\nu,\nu} }
u(B,q)_{i,\nu} U_{i,\nu} \delta_{i,j-\hat\nu}
\right. \nonumber \\ && - \left.
u^*(B,q)_{i - \hat\nu,\nu} U^{\dag}_{i-\hat\nu,\nu}\delta_{i,j+\hat\nu} 
\right) \:.
\label{fmatrix1}
\end{eqnarray}
$\mathcal{D}U$ is the functional integration over the gauge link
variables $U_{n,\mu}$, $S_G$ is the discretized pure gauge action (we consider a
standard Wilson action).  The subscripts $i$ and $j$ refer to lattice
sites, $\hat\nu$ is a unit vector on the lattice and $\eta_{i,\nu}$ are the staggered
phases. Periodic (antiperiodic) boundary conditions (b.c.) must be taken for gauge
(fermion) fields along the Euclidean time direction, while spatial 
periodic b.c. are chosen for all fields. $u(B,q)_{i,\nu}$ are the
gauge links corresponding to the background $U(1)$ magnetic field.  We shall consider
a constant magnetic field $\vec B = B \hat z$ and 
the following choice for the gauge field:
\beq
A_y = B x
\:; \qquad\qquad
A_\mu = 0
\quad \mathrm{for} \quad \mu = x,z,t
\:.
\eeq
The corresponding $U(1)$ links on the lattice are:
\beq
u(B,q)_{n,y} = e^{i a^2 q B n_x}
\, ; \quad
u(B,q)_{n,\mu} = 1
\:\: \mathrm{for} \:\: \mu = x,z,t\
\label{u1field}
\eeq
This choice corresponds to a magnetic flux $a^2 B$ going through each plaquette in
the $x-y$ plane, except at the boundary $(L_x,y,z,t)$, due to  the
periodic b.c. in the spatial directions. In order to guarantee the
smoothness of the background field across the boundary and the gauge invariance of
the fermion action the $U(1)$ gauge fields must be modified at the boundary of the
$x$ direction:
\beq
u(B,q)_{n,x=L_x} = e^{-i a^2 q L_x B n_y}
\label{boundary}
\:
\eeq
and the magnetic field must be quantized, $a^2qB=2\pi b/L_xL_y$, where $b$ is an integer. 
That corresponds to taking the appropriate gauge 
invariant b.c. for fermion fields on the torus~\cite{wiese} 
(with the possible additional free phases
$\theta_x$ and $\theta_y$~\cite{wiese} set to zero).
Given the two different values of $q_u$ and $q_d$, the quantization of $B$
in our case is set by the $d$ quark charge $q_d = -|e|/3$,
\beq
|e| B = 6 \pi T^2 \left(\frac{N_t}{L_s}\right)^2 b
\:,
\eeq
$T = 1/(N_t a)$ is the temperature and $L_x = L_y \equiv L_s$.

Our simulations have been carried out on $16^3 \times 4$ lattices and for
three different bare quark masses $am = 0.01335$, $0.025$ and
$0.075$. The corresponding (Goldstone) pion masses are $am_\pi = 0.307(3), 0.417(3)$
and $0.707(3)$. The temperature $T = 1/(N_t a)$ is changed by varying the lattice
spacing through the inverse gauge coupling $\beta$.

Zero $T$ estimates of the string tension, done at the same $\beta$ values where
the $B = 0$ transition takes place, lead to $a$ ranging from 0.29
to 0.31 fm as $am$ is decreased, corresponding to $T_c(B = 0)$ ranging from
170 to 160 MeV.  
The corresponding values of the (Goldstone) pion
mass are $m_\pi\approx 195, 275$ and $480$ MeV. For each quark mass we have
done simulations using magnetic field corresponding to $b=0,8,16$ and $24$, {\it
i.e.} for $|e|B=0,3\pi T^2, 6\pi T^2$ and $9\pi T^2$ respectively. Thus, for the
lightest pion mass
our magnetic field reaches values up to $|e|B\approx19\ m_\pi^2$, i.e.
$\sqrt{|e|B}\approx850$ MeV in physical units. Note that, since we
are working with a fixed value of $N_t$, $B$ changes while changing
temperature as $a^{-2}\propto T^2$. However, for all the quark masses the range of
couplings (hence of $a$) that we explore corresponds to a $<2\%$ change
in $T$ and hence the magnetic field only changes at most by a few percent. 

The Rational Hybrid Monte-Carlo algorithm has been used to simulate 
rooted staggered fermions: we need to treat separately each flavor and thus take the
fourth root of the fermion determinant. Typical statistics are of the order of 10k
molecular dynamics trajectories.

\section{Numerical Results}
\label{results}

\begin{figure}[h!]
\includegraphics*[width=0.47\textwidth]{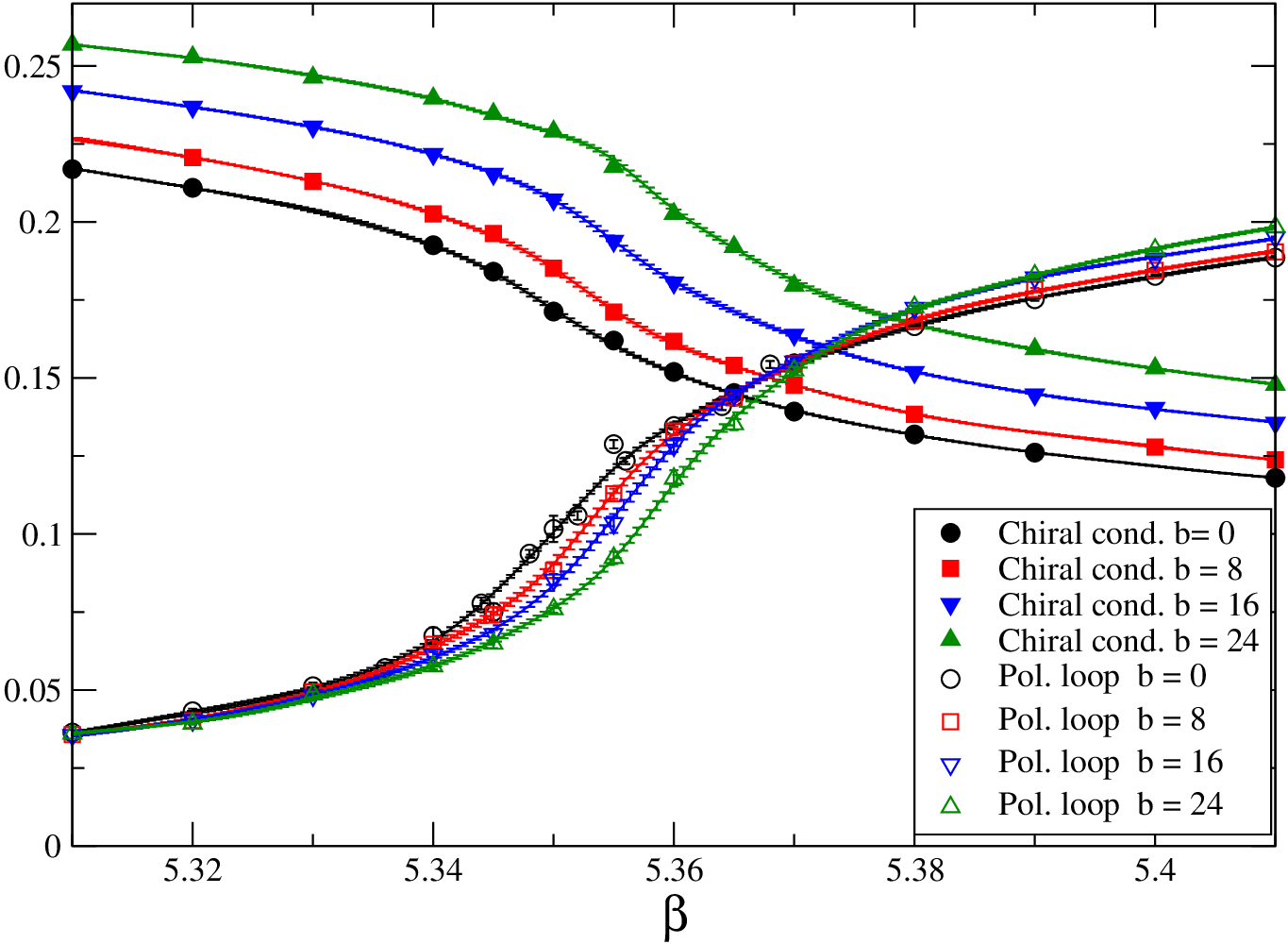}
\caption{Chiral condensate and Polyakov loop for $a m = 0.075$}
\label{chiralpol0.075}
\end{figure}

\begin{figure}[h!]
\includegraphics*[width=0.47\textwidth]{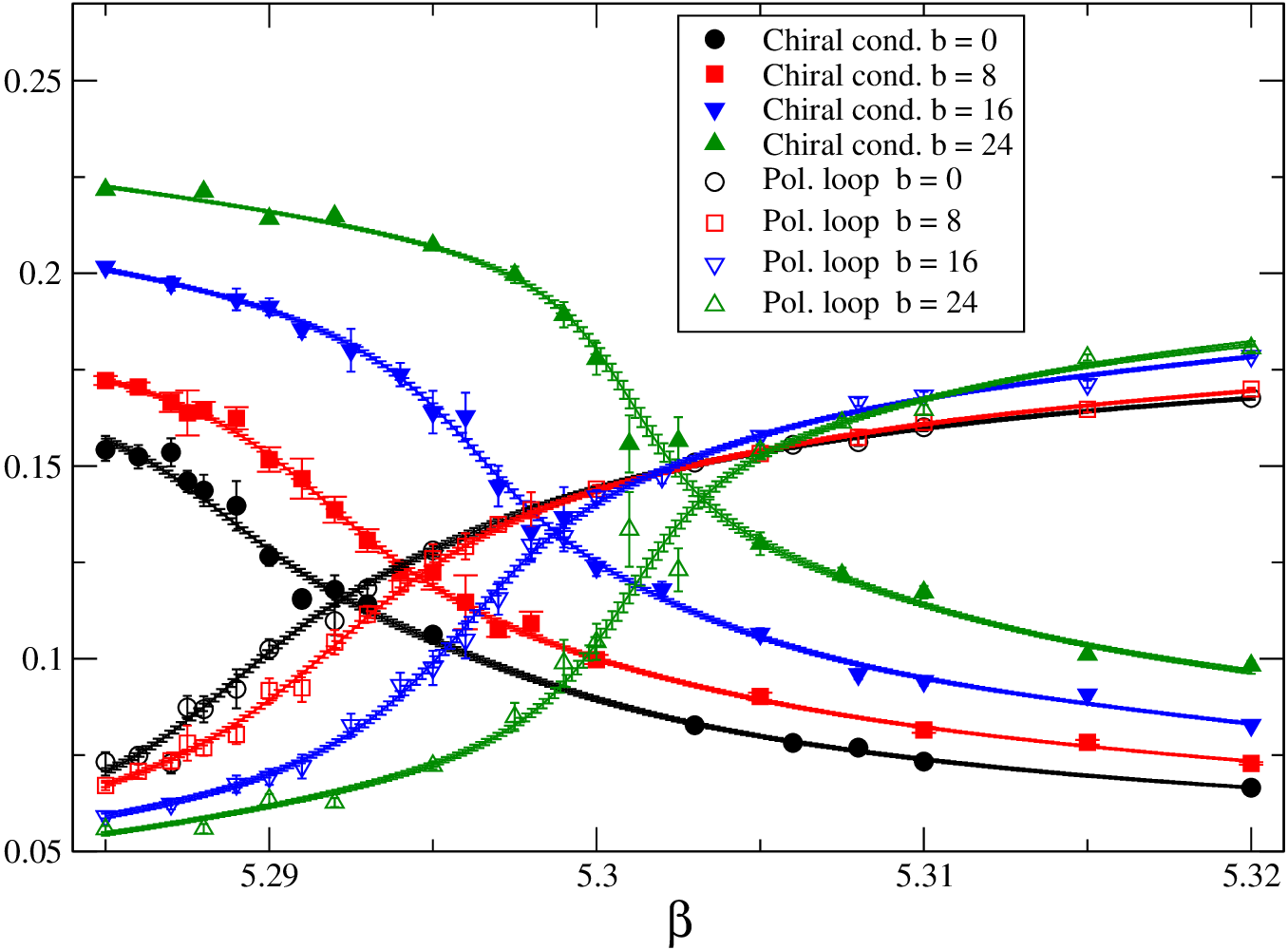}
\caption{Same as in Fig.~\ref{chiralpol0.075} for $a m = 0.025$}
\label{chiralpol0.025}
\end{figure}

\begin{figure}[h!]
\includegraphics*[width=0.47\textwidth]{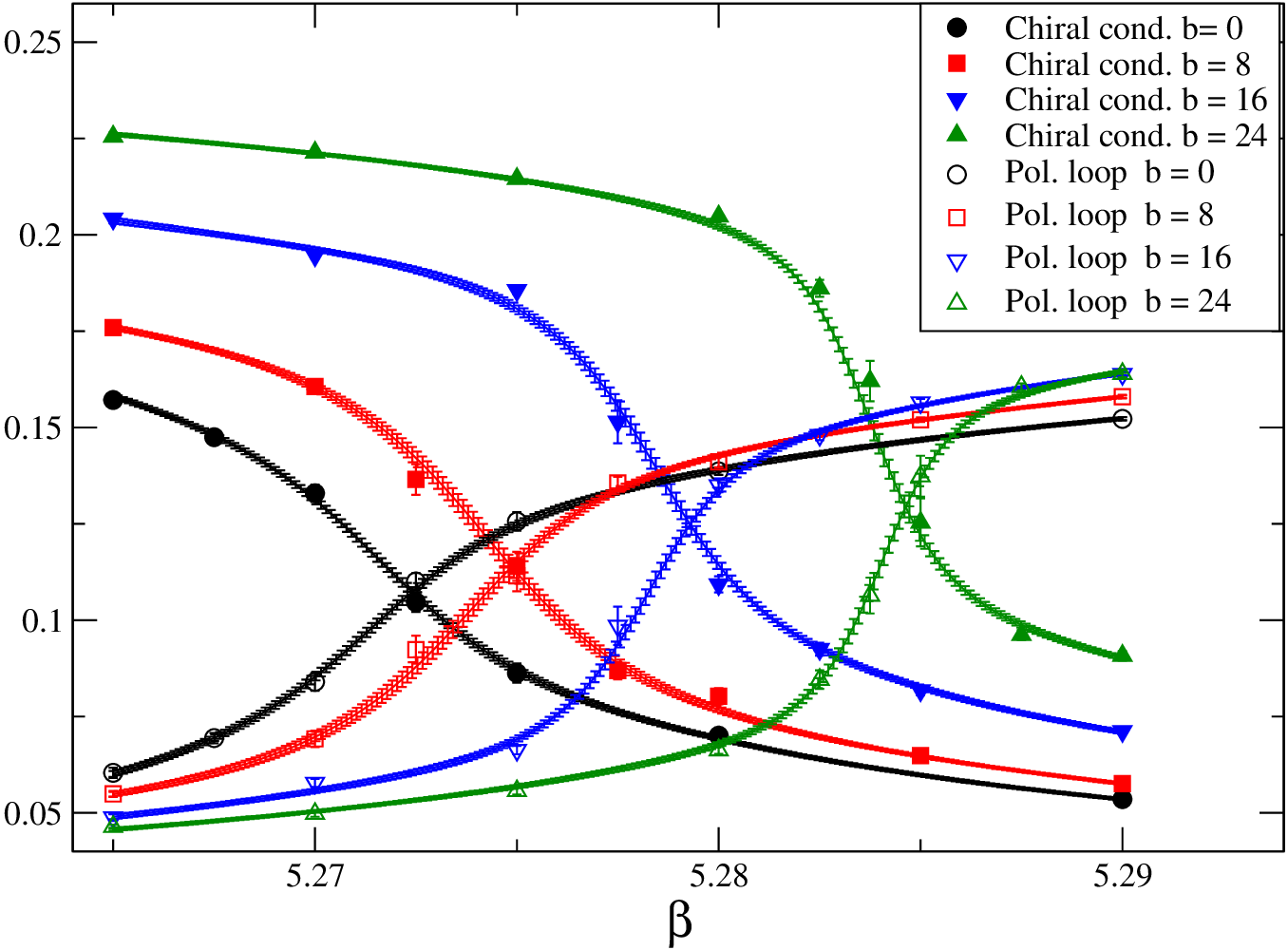}
\caption{Same as in Fig.~\ref{chiralpol0.075} for $a m = 0.01335$}
\label{chiralpol0.01335}
\end{figure}

\begin{figure}[h!]
\includegraphics*[width=0.47\textwidth]{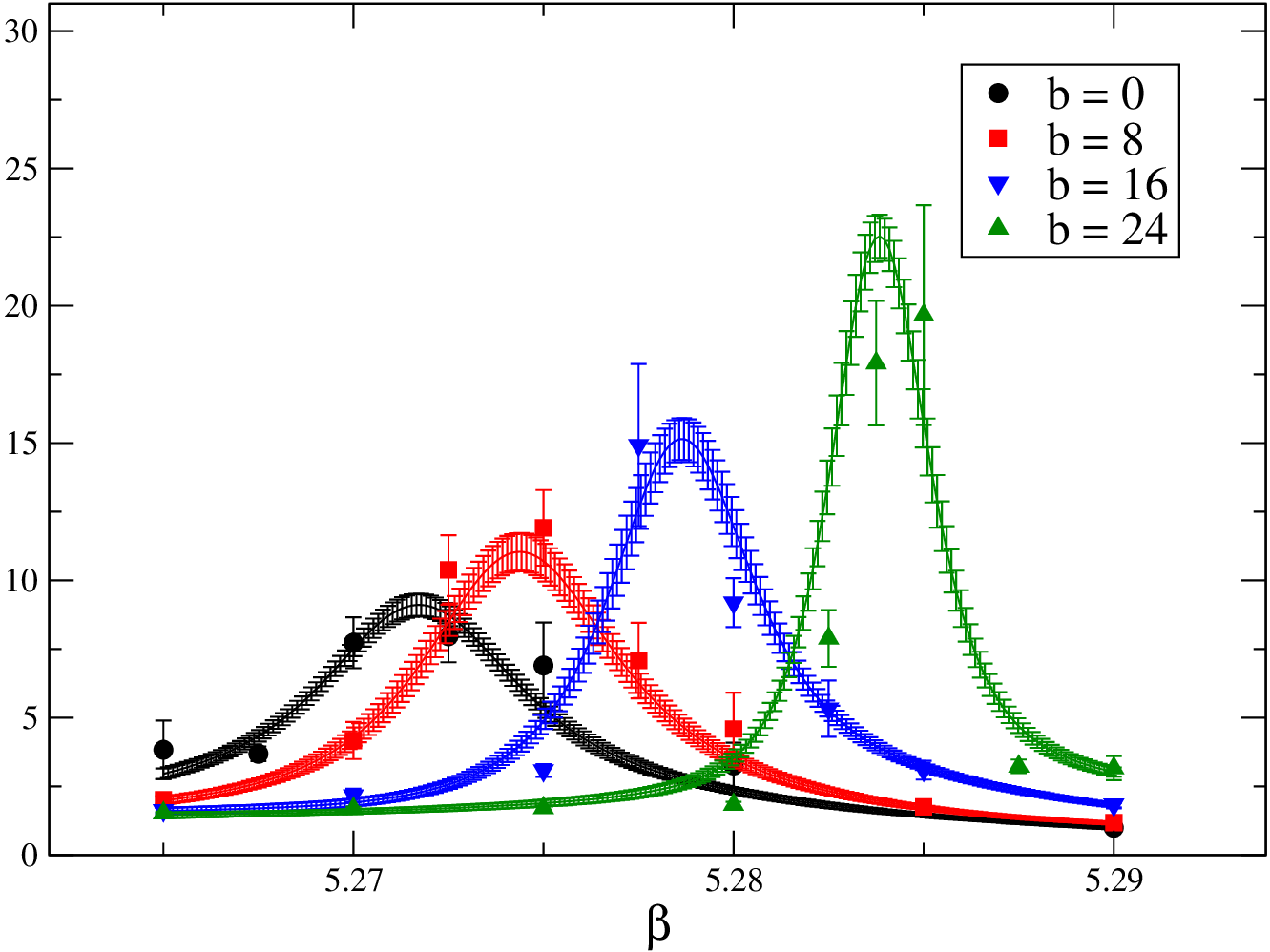}
\caption{Disconnected $\langle\bar\psi\psi\rangle$ susceptibility 
for $a m = 0.01335$}
\label{suscchiral0.01335}
\end{figure}

In Figs.~\ref{chiralpol0.075}, \ref{chiralpol0.025} and \ref{chiralpol0.01335}
we show the behavior of $\langle\bar\psi\psi\rangle$
(average of the $u$ and $d$ quark condensates)
and of the Polyakov loop for different magnetic fields and $am = 0.075$, $0.025$ 
and $0.01335$. 
Results are presented as a function of the inverse gauge coupling $\beta$:
we recall that $T$ is an increasing function of 
$\beta$, a conversion into physical units will be presented later.
For all temperatures $\langle\bar\psi\psi\rangle$
increases as a function of $B$,
as expected from analytic predictions. Interpreting the drop of the condensate as 
the signal for chiral symmetry restoration, we infer that the transition temperature increases
as a function of $B$. 
A sharper
drop is observed at the highest fields explored, especially for the lowest quark masses, 
indicating a sizable increase in the 
strength of the transition; that is 
visible from the behavior of the disconnected
chiral susceptibility (Fig.~\ref{suscchiral0.01335}).

\begin{figure}[h!]
\includegraphics*[width=0.47\textwidth]{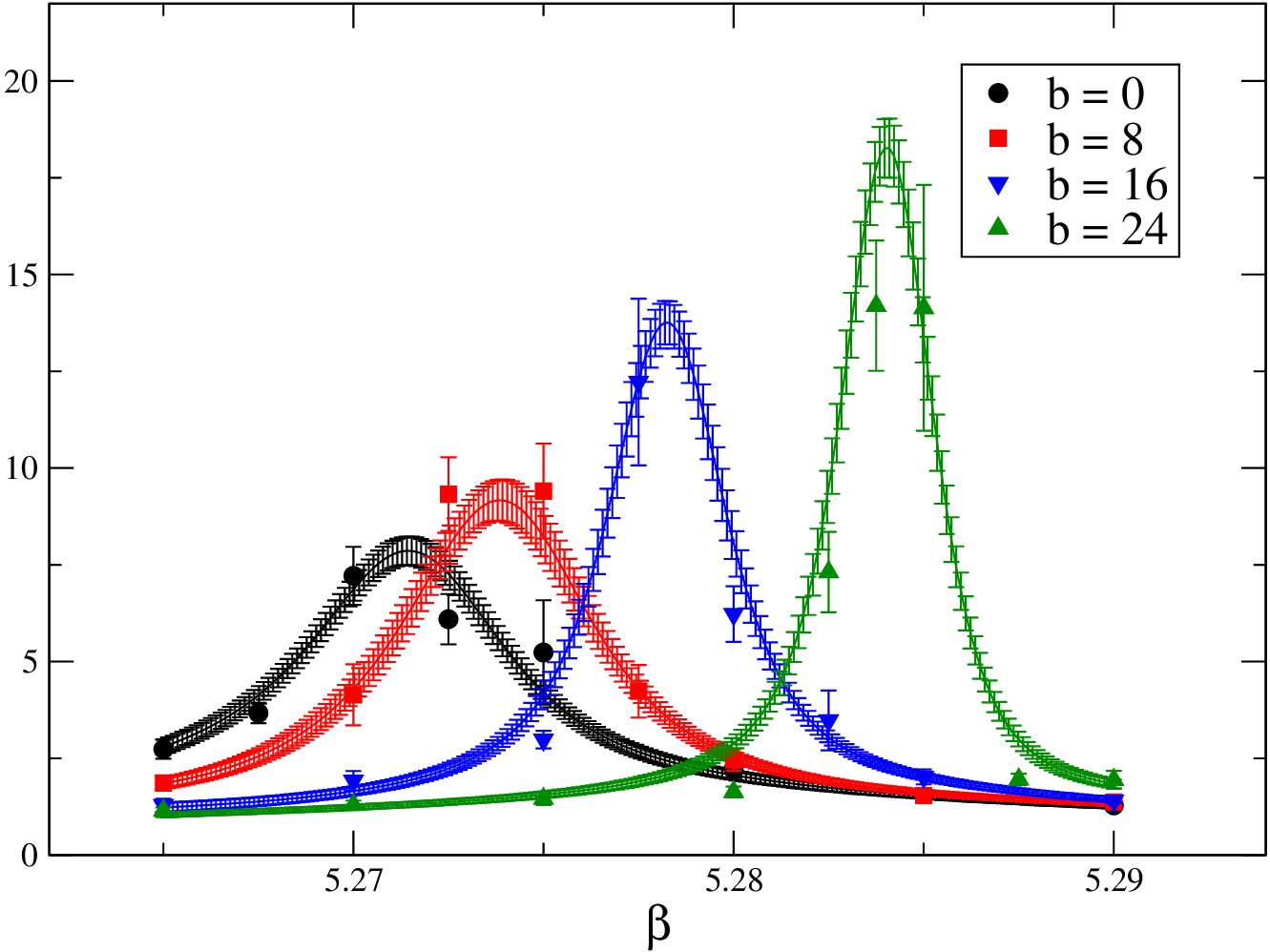}
\caption{Polyakov loop susceptibility for $a m = 0.01335$}
\label{suscpol0.01335}
\end{figure}

The Polyakov loop $P$ is a pure gauge quantity, coupled to the magnetic field only through
quark loops, 
hence its behavior is less trivial to predict.
From Figs.~\ref{chiralpol0.075}, \ref{chiralpol0.025} and \ref{chiralpol0.01335} we see that, while
at low T it decreases as a function of $B$ (as one would expect
qualitatively from the fact that $\langle\bar\psi\psi\rangle$ increases), at high $T$ it increases. 
Such behavior is 
qualitatively similar to what obtained in 
Ref.~\cite{fukushima} by a PNJL model analysis and should be further
investigated, e.g. by determining the renormalized Polyakov loop. 
If we interpret the rise of $P$ as the onset of  deconfinement, we infer that
the shift and the increase in strength of the deconfinement transition is similar to what
observed for the chiral transition. Data obtained for the Polyakov loop susceptibility
lead to similar conclusions (see Fig.~\ref{suscpol0.01335}). 

\begin{table}
\begin{center}
\begin{tabular}{|c|c|c|c|}
\hline
$a m_q$ & $b$ & $\beta_c$ (Pol. loop) & $\beta_c$ ($\bar \psi \psi$)  \\ \hline
0.01335 & 0  & 5.2714(4)  & 5.2716(3)\\
0.01335 & 8  & 5.2739(4)  & 5.2741(4)\\
0.01335 & 16 & 5.2783(3)  & 5.2785(3)\\
0.01335 & 24 & 5.2836(2)  & 5.2838(2)\\ \hline
0.025  & 0   & 5.2893(2)  & 5.2898(3)\\
0.025  & 8   & 5.2925(3)  & 5.2925(3)\\
0.025  & 16  & 5.2961(3)  & 5.2966(3)\\
0.025  & 24  & 5.3014(4)  & 5.3018(4)\\ \hline
0.075  & 0   & 5.351(1)  & 5.351(2)\\
0.075  & 8   & 5.353(1)  & 5.353(2)\\
0.075  & 16  & 5.355(1)  & 5.357(2)\\
0.075  & 24  & 5.358(1)  & 5.360(1)\\ \hline
\end{tabular}
\end{center}
\caption{Pseudocritical couplings obtained by fitting the peak of the 
chiral condensate or Polyakov loop susceptibilities.}
\label{couplings}
\end{table}

In Table~\ref{couplings} we report the pseudocritical couplings $\beta_c$ 
for deconfinement and chiral
restoration obtained by fitting the peak of the susceptibilities by a quadratic function.
We have also determined $\beta_c$ looking for the inflection point of observables, by means
of polynomial fits, obtaining compatible results within errors.
Data obtained for $\beta_c$ confirm that 
no 
appreciable separation of chiral restoration and deconfinement is induced by the 
the background field, at least for the explored field strengths.

\begin{figure}[h!]
\includegraphics*[width=0.47\textwidth]{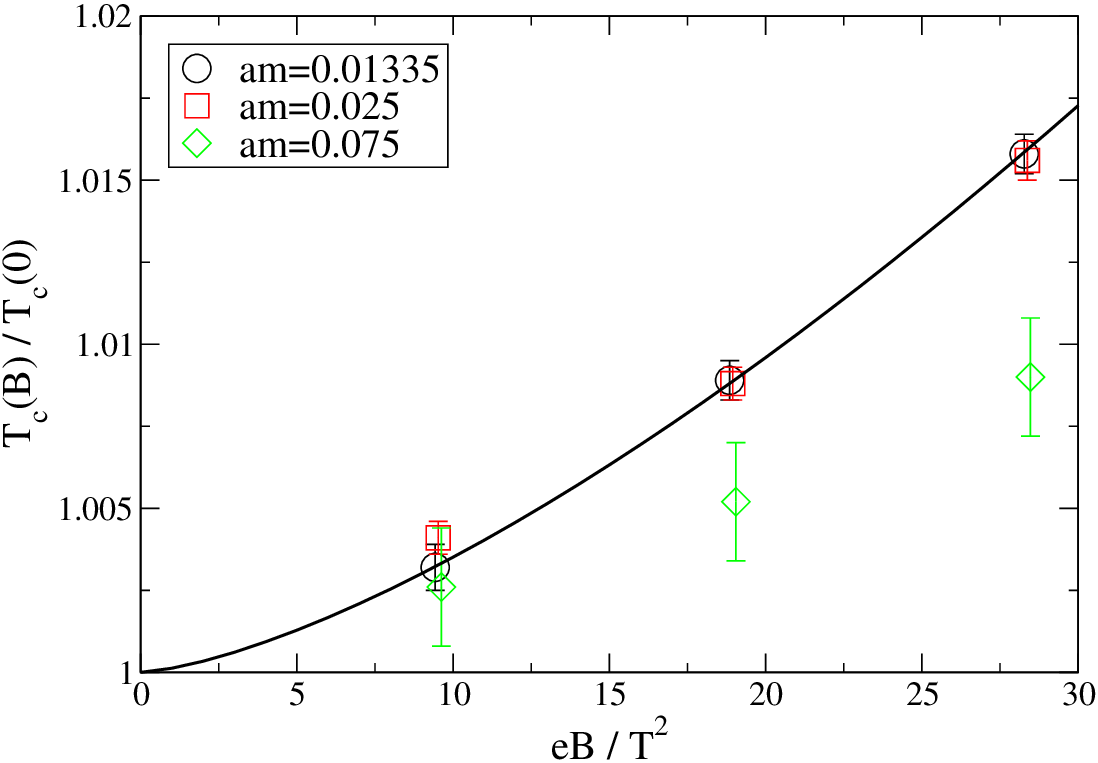}
\caption{$T_c(B)$ for different quark masses. The solid curve is 
a power law fit to the lightest quark data (see text).}
\label{ttcfigure}
\end{figure}

From the values of $\beta_c$ we obtain the ratio $T_c (B) / T_c(0)$ as 
a function of the dimensionless ratio $eB/T^2$, as reported in Fig.~\ref{ttcfigure}; 
the 2-loop $\beta$-function has been used for the conversion.
A direct determination of the physical scale on $T = 0$ lattices
is preferable but would require very precise measurements to appreciate
$T$ variations of the order of percent. On the other hand, 
given the small scale range explored, 
the approximation is reasonable and no qualitative
change is expected. 

\begin{figure}[h!]
\includegraphics*[width=0.47\textwidth]{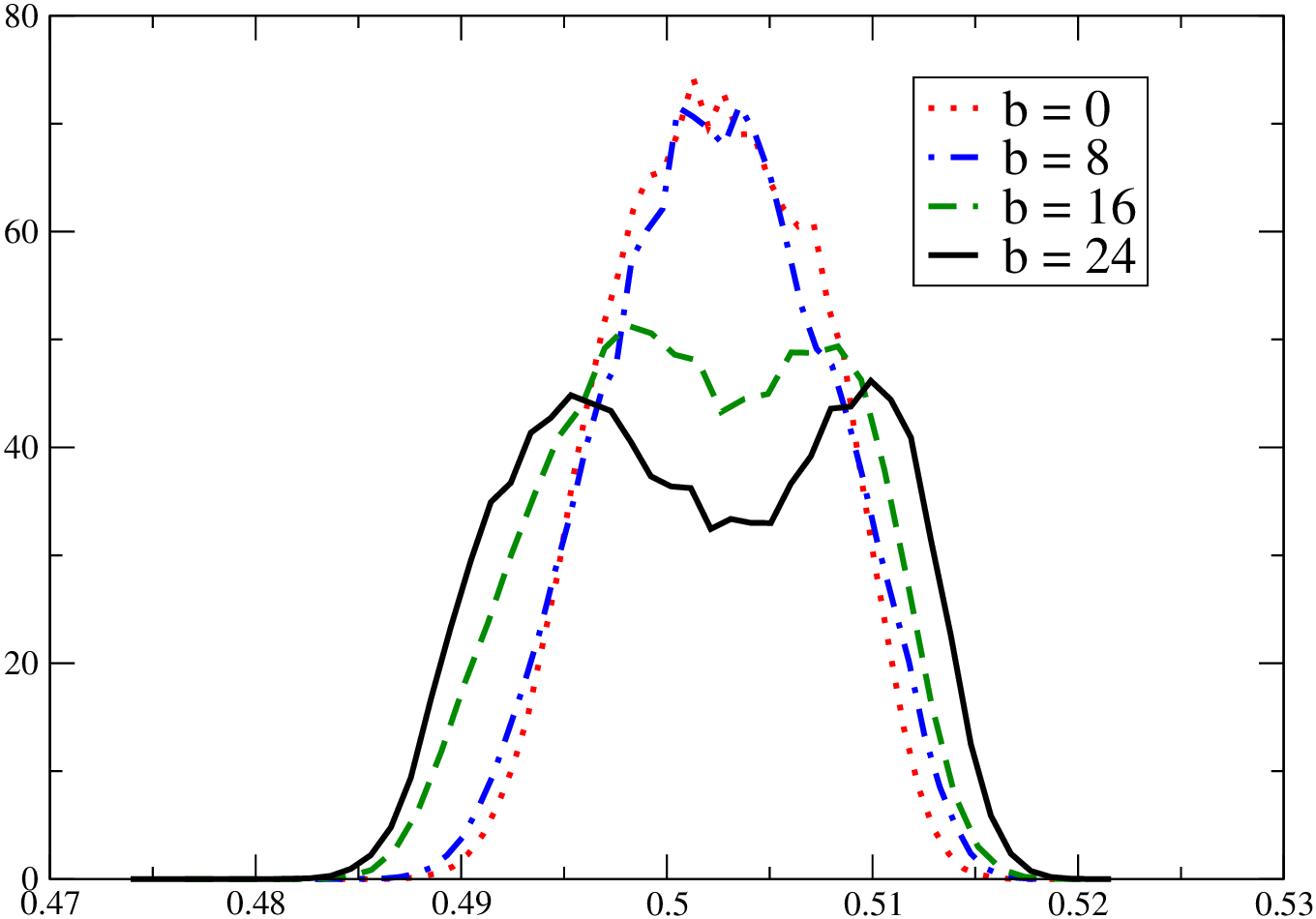}
\caption{Reweighted plaquette distribution at $\beta_c$ as a function of the external field
at $am = 0.01335$ on a $16^3\times 4 $ lattice.} 
\label{histoplaq16_0.01335}
\end{figure}

\begin{figure}[h!]
\includegraphics*[width=0.47\textwidth]{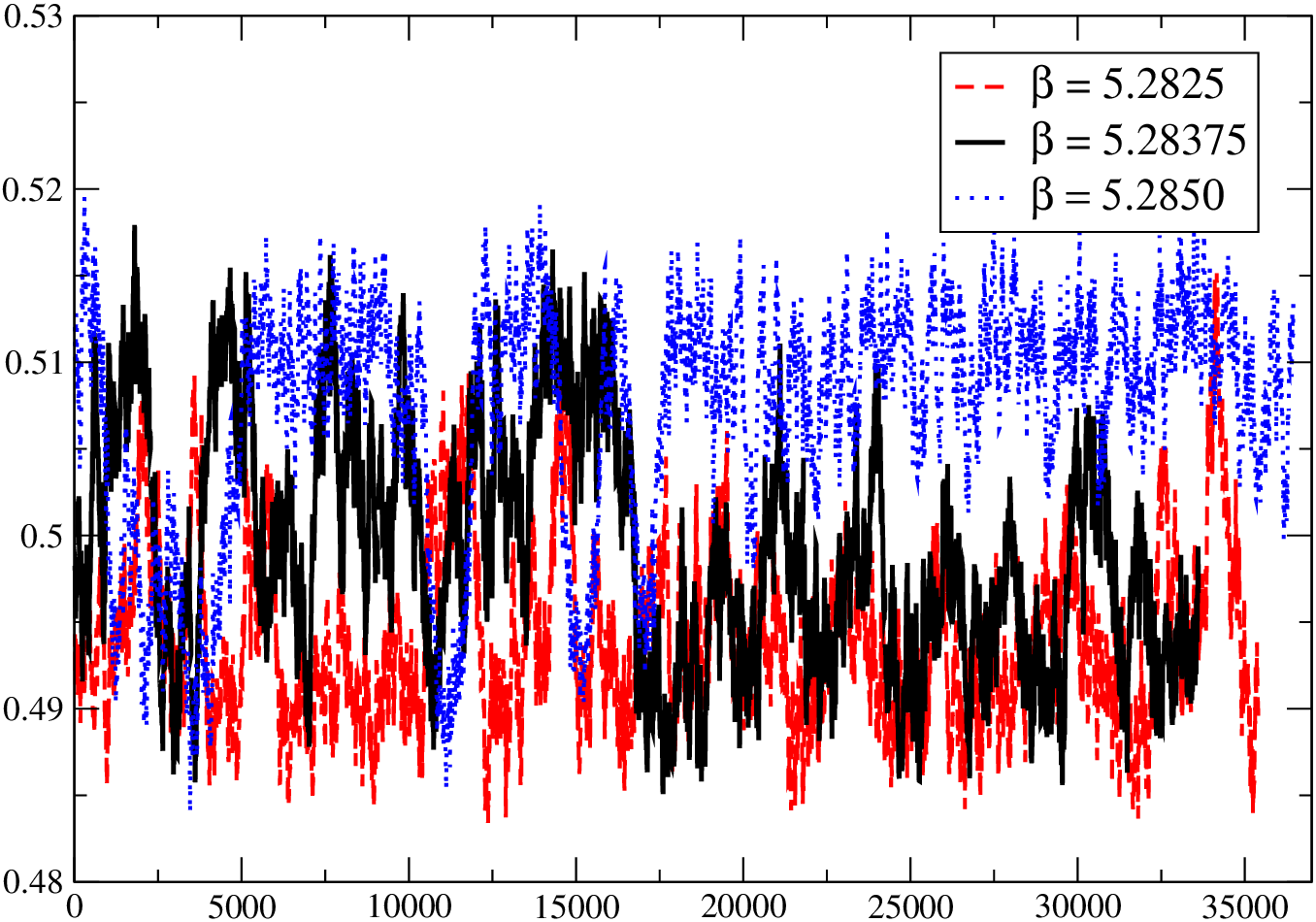}
\caption{Monte-Carlo histories of the plaquette at 3 different $\beta$ values around the transition for $b = 24$ and $am = 0.01335$.} 
\label{histoplaq16_0.01335}
\end{figure}

Fig.~\ref{ttcfigure} shows that the change in $T_c$ is small
and of the order of a few percent at the highest explored fields. Moreover, there seems
to be a saturation as the chiral limit is approached: results for $am = 0.01335$ and
$am = 0.025$
stay onto each other. Notice that this is true if we plot results as a function 
of $|e|B/T^2$: had we used $|e|B/m_\pi^2$ results at different masses would have been 
different: the highest $B$ is about 20 $m_\pi^2$ for $am = 0.01335$ 
and 10 $m_\pi^2$ for $am = 0.025$. This suggests that, at
least for the strong fields and for the pion masses explored, 
the relevant scale governing the effect of the magnetic field on the shift 
of the transition 
is $T$ itself and not $m_\pi$.

Trying to understand the dependence of $T_c$ on $B$,
we have  fitted our data for $am = 0.01335$ according to 
\beq
\frac{T_c(B)}{T_c(0)} = 1 + A \left( \frac{|e|B}{T^2} \right)^\alpha
\eeq
finding that $\alpha = 1.45(20)$ and $A \sim 1.3\ 10^{-4}$.

Finally we discuss about the nature of the transition. 
At $B = 0$ it is still unclear if a weak first order transition is present
in the chiral limit~\cite{pisa1,pisa2}, however no clear
signal for a finite latent heat has been found
on available lattice sizes, hence the first order transition, even if present, is so
weak to be of poor phenomenological relevance.
On the other hand our results show that the introduction of a magnetic field makes
the transition sharper.
The question is if large fields can turn the transition into a 
first order strong enough to be clearly detectable.

To that aim we have analyzed the reweighted 
plaquette distribution at the critical couplings and for different values of $B$: 
results are shown in Fig.~\ref{histoplaq16_0.01335}. 
The single peak distribution, which is present at zero or small magnetic field, turns
into a double peak distribution, typical of a first order transition,
for the largest $B$ explored; also the Monte-Carlo
histories of the plaquette, Fig.~\ref{histoplaq16_0.01335}, 
present signals of a metastable behavior.
 We can consider that as an indication but 
not as a final answer: numerical simulations on larger lattice sizes are necessary
to clarify if the double peak structure survives the thermodynamical limit and for 
a proper finite size scaling analysis.

\section{Conclusions}
\label{conclusions}

We have presented results from an investigation of the $N_f = 2$ QCD phase diagram in presence 
of a magnetic background field.
We have explored different quark masses,
corresponding to $m_\pi$ ranging from 200 MeV to 480 MeV, and different  
magnetic fields, with $\sqrt{|e|B}$ up to about 850 MeV ($|e| B \sim 20\ m_\pi^2$ for the 
lightest mass).

Main results can be summarized as follows: the transition temperature increases slightly 
( $ <$ 2\% at the highest field) and no evidence is found, 
within the range of explored fields, for a disentanglement of chiral symmetry 
restoration and deconfinement. $T_c(B)/T_c(0)$ as a function of $|e|B/T^2$ shows negligible 
dependence on $m_\pi$ for the two lowest masses, and is well described by a power law
$T_c(B)/T_c(0) = 1 + A (|e|B/T^2)^\alpha$ with $\alpha \sim 1.45(20)$.
The transition becomes sharper with some
preliminary evidence for a first order transition, in the form of double peak distributions,
at the highest fields explored: such indications should be clarified by future studies on 
larger spatial volumes and by a finite size scaling analysis.

Regarding the comparison with model predictions, 
our results show partial agreement with some of the results reported in 
Ref.~\cite{fraga2} and in Ref.~\cite{fukushima}: the deconfinement and chiral restoring
temperatures both increase, even if we do not see any sign for a faster
grow and splitting of the chiral transition till $|e|B \sim 20$ $m_\pi^2$.
Also the observed increase in the strength of the transition is common
to some models~\cite{fraga1,fraga2}. 
We stress the qualitatively different behavior which is observed in 
numerical simulations with a background chromo-magnetic field, where
$T_c$ decreases as a function of the external field~\cite{bari1,bari2}.

Our results have been obtained using a standard staggered 
discretization and a coarse lattice, with a lattice spacing $a \sim 0.3$ fm. Apart
from possible systematic effects related to the fourth root trick, flavour breaking 
discretization effects may play an important role, with a distorted hadron spectrum that
could partially modify the effect of the magnetic field on QCD thermodynamics. For this 
reason it will be important to confirm our results in the future by using
different lattice discretizations.

\begin{acknowledgements}
Numerical simulations have been carried out on two computer farms in Genova and Bari
and on the apeNEXT facilities in Rome. We thank E. Fraga, K. Fukushima, K. Klimenko, M. Ruggieri and H. Suganuma for useful discussions. 
SM has been supported by
contract DE-AC02-98CH10886 with the U.S. Department of Energy.
\end{acknowledgements}


\begin{thebibliography}{9}

\bibitem{Vachaspati:1991nm}
  T.~Vachaspati,
  Phys.\ Lett.\  B {\bf 265}, 258 (1991).

\bibitem{heavyionfield1}
  D.~E.~Kharzeev, L.~D.~McLerran and H.~J.~Warringa,
  Nucl.\ Phys.\  A {\bf 803}, 227 (2008).

\bibitem{heavyionfield2}
  V.~Skokov, A.~Y.~Illarionov and V.~Toneev,
  Int.\ J.\ Mod.\ Phys.\  A {\bf 24}, 5925 (2009).

\bibitem{magnetars}
  R.~C.~Duncan and C.~Thompson,
  Astrophys.\ J.\  {\bf 392}, L9 (1992).

\bibitem{Klevansky:1989vi}
  S.~P.~Klevansky and R.~H.~Lemmer,
  Phys.\ Rev.\  D {\bf 39}, 3478 (1989).

\bibitem{Suganuma}
  H.~Suganuma and T.~Tatsumi,
  Annals Phys.\  {\bf 208}, 470 (1991).

\bibitem{Gusynin:1994xp}
  V.~P.~Gusynin, V.~A.~Miransky and I.~A.~Shovkovy,
  Phys.\ Lett.\  B {\bf 349}, 477 (1995).

\bibitem{Shushpanov:1997sf}
  I.~A.~Shushpanov and A.~V.~Smilga,
  Phys.\ Lett.\  B {\bf 402}, 351 (1997).

\bibitem{Ebert}
  D.~Ebert, K.~G.~Klimenko, M.~A.~Vdovichenko and A.~S.~Vshivtsev,
  Phys.\ Rev.\  D {\bf 61}, 025005 (2000).

\bibitem{Miransky:2002rp}
  V.~A.~Miransky and I.~A.~Shovkovy,
  Phys.\ Rev.\  D {\bf 66}, 045006 (2002).

\bibitem{Cohen:2007bt}
  T.~D.~Cohen, D.~A.~McGady and E.~S.~Werbos,
  Phys.\ Rev.\  C {\bf 76}, 055201 (2007).

\bibitem{Zayakin:2008cy}
  A.~V.~Zayakin,
  JHEP {\bf 0807}, 116 (2008).

\bibitem{cme0}
  D.~E.~Kharzeev, L.~D.~McLerran and H.~J.~Warringa,
  Nucl.\ Phys.\  A {\bf 803}, 227 (2008).

\bibitem{cme1}
  K.~Fukushima, D.~E.~Kharzeev and H.~J.~Warringa,
  Phys.\ Rev.\  D {\bf 78}, 074033 (2008).

\bibitem{star}
  B.~I.~Abelev {\it et al.}  [STAR Collaboration],
  Phys.\ Rev.\ Lett.\  {\bf 103}, 251601 (2009).

\bibitem{Klimenko}
  K.~G.~Klimenko,
  Z.\ Phys.\  C {\bf 54}, 323 (1992).

\bibitem{agasian}
  N.~O.~Agasian and S.~M.~Fedorov,
  Phys.\ Lett.\  B {\bf 663}, 445 (2008).

\bibitem{fraga1}
  E.~S.~Fraga and A.~J.~Mizher,
  Phys.\ Rev.\  D {\bf 78}, 025016 (2008);

\bibitem{NJL}
  J.~K.~Boomsma and D.~Boer,
  Phys.\ Rev.\  D {\bf 81}, 074005 (2010).

\bibitem{fukushima}
  K.~Fukushima, M.~Ruggieri and R.~Gatto,
  arXiv:1003.0047 [hep-ph].

\bibitem{fraga2}
  A.~J.~Mizher, M.~N.~Chernodub and E.~S.~Fraga,
  arXiv:1004.2712 [hep-ph].

\bibitem{gatto}
  R.~Gatto and M.~Ruggieri,
  arXiv:1007.0790 [hep-ph].


\bibitem{bari1}
P.~Cea and L.~Cosmai,
  JHEP {\bf 0508}, 079 (2005).

\bibitem{bari2}
  P.~Cea, L.~Cosmai and M.~D'Elia,
  JHEP {\bf 0712}, 097 (2007).

\bibitem{hadron1}
  G.~Martinelli, G.~Parisi, R.~Petronzio and F.~Rapuano,
  Phys.\ Lett.\  B {\bf 116}, 434 (1982).

\bibitem{hadron2}
  C.~W.~Bernard, T.~Draper, K.~Olynyk and M.~Rushton,
  Phys.\ Rev.\ Lett.\  {\bf 49}, 1076 (1982).

\bibitem{cme-itep}
  P.~V.~Buividovich, M.~N.~Chernodub, E.~V.~Luschevskaya and M.~I.~Polikarpov,
  Phys.\ Rev.\  D {\bf 80}, 054503 (2009).

\bibitem{cme-blum}
  M.~Abramczyk, T.~Blum, G.~Petropoulos and R.~Zhou,
  arXiv:0911.1348 [hep-lat].

\bibitem{itep1}
  P.~V.~Buividovich, M.~N.~Chernodub, E.~V.~Luschevskaya and M.~I.~Polikarpov,
  Phys.\ Lett.\  B {\bf 682}, 484 (2010),
  Nucl.\ Phys.\  B {\bf 826}, 313 (2010).

\bibitem{kharzeev-itep}
  P.~V.~Buividovich {\it et al.},
  arXiv:1003.2180 [hep-lat].

\bibitem{wiese}
  M.~H.~Al-Hashimi and U.~J.~Wiese,
  Annals Phys.\  {\bf 324}, 343 (2009).

\bibitem{pisa1}
  M.~D'Elia, A.~Di Giacomo and C.~Pica,
  Phys.\ Rev.\  D {\bf 72}, 114510 (2005).

\bibitem{pisa2}
  G.~Cossu {\it et al.}, 
  arXiv:0706.4470 [hep-lat].

\end{thebibliography}
\end{document}